\documentclass[a4paper,11pt]{article}
\pdfoutput=1 
\usepackage{jinstpub} 
\usepackage{hyperref}
\usepackage{lineno,hyperref}
\usepackage{amsmath}
\usepackage[utf8]{inputenc} 

\title{\boldmath Commissioning and testing of pre-series triple GEM prototypes for CBM-MuCh in the mCBM experiment at the SIS18 facility of GSI}

\author[a,b,1]{A.~Kumar, \note{Corresponding author.}}
\author[a,b]{A.~Agarwal,}
\author[c]{S.~Chatterjee,}
\author[a,b]{S.~Chattopadhyay,}
\author[a,b]{A.~K.~Dubey,}
\author[a,b]{C.~Ghosh, }
\author[a,b]{E.~Nandy,}
\author[a]{V.~Negi,}
\author[c]{S.~K.~Prasad,}
\author[a]{J.~Saini,}
\author[a,b]{V.~Singhal,}
\author[e]{O.~Singh,}
\author[d]{G.~Sikder,}
\author[g]{J.~de~Cuveland,}
\author[h]{I.~Deppner,}
\author[f]{D.~Emschermann,}
\author[f]{V.~Friese,}
\author[f]{J.~Fr\"uhauf,}
\author[i]{M.~Gumiński,}
\author[h]{N.~Herrmann,}
\author[g]{D.~Hutter,}
\author[f]{M.~Kis,}
\author[f]{J.~Lehnert,}
\author[f]{P.-A.~Loizeau,}
\author[f]{C.J.~Schmidt,}
\author[f]{C.~Sturm,}
\author[f]{F.~Uhlig,}
\author[i]{W.~Zabołotny}

\affiliation[a]{Variable Energy Cyclotron Centre, Kolkata, INDIA}
\affiliation[b]{Homi Bhabha National Institute, Mumbai, INDIA}
\affiliation[c]{Bose Institute, Kolkata, INDIA}
\affiliation[d]{University of Calcutta, Kolkata, INDIA}
\affiliation[e]{Aligarh Muslim University, Aligarh, India}
\affiliation[f]{GSI Helmholtz Center for Heavy-Ion Research GmbH (GSI), Darmstadt, Germany}
\affiliation[g]{Frankfurt Institute of Advanced Studies (FIAS), Frankfurt am Main, Germany}
\affiliation[h]{Physikalisches Institut, Ruprecht-Karls-Universit{\"a}t Heidelberg, Germany}
\affiliation[i]{Institute of Electronic Systems, Warsaw University of Technology, Warsaw, Poland}

\emailAdd{akmaurya@vecc.gov.in}

\abstract{Large area triple GEM chambers will be employed in the first two stations of the MuCh system of the CBM experiment at the upcoming Facility for Antiproton and Ion Research FAIR in Darmstadt/Germany. The GEM detectors have been designed to take data at an unprecedented interaction rate (up to 10 MHz) in nucleus-nucleus collisions in CBM at FAIR. Real-size trapezoidal modules have been installed in the mCBM experiment and tested in nucleus-nucleus collisions at the SIS18 beamline of GSI as a part of the FAIR Phase-0 program. In this report, we discuss the design, installation, commissioning, and response of these GEM modules in detail. The response has been studied using the free-streaming readout electronics designed for the CBM-MuCh and CBM-STS detector system. In free-streaming data, the first attempt on an event building based on the timestamps of hits has been carried out, resulting in the observation of clear spatial correlations between the GEM modules in the mCBM setup for the first time. Accordingly, a time resolution of $\sim$15\,ns have been obtained for the GEM detectors.}


\begin{document}
\flushbottom
\maketitle

\linenumbers

\section{Introduction}
\label{sec:intro}
The Compressed Baryonic Matter (CBM) experiment~\cite{Cbm} at the upcoming Facility for Antiproton and Ion Research (FAIR) in Darmstadt, Germany~\cite{Fair} is designed to investigate the properties of dense nuclear matter in nucleus-nucleus collision at an unprecedented interaction rate up to 10 MHz, which allows us to study extremely rare probes such as low mass vector mesons (LMVM), multi-strange hadrons and charmonia. The energy of the colliding beams will vary from 2-29 AGeV for proton and 2-14 AGeV for heavy-ion beams in the SIS100 ring of FAIR.

\begin{figure}[htbp]
\centering 
\includegraphics[width=5.2cm,height=5.5cm]{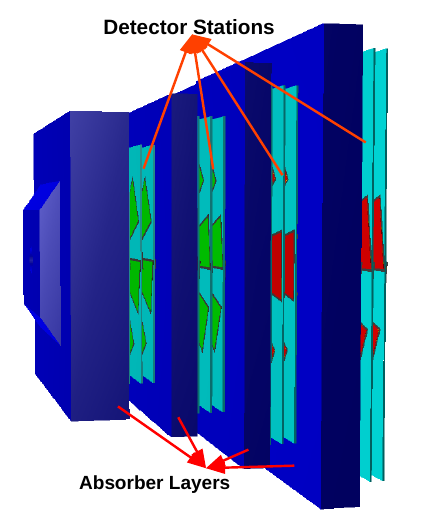}
\hspace{1mm}
\includegraphics[width=4.0cm,height=5.5cm]{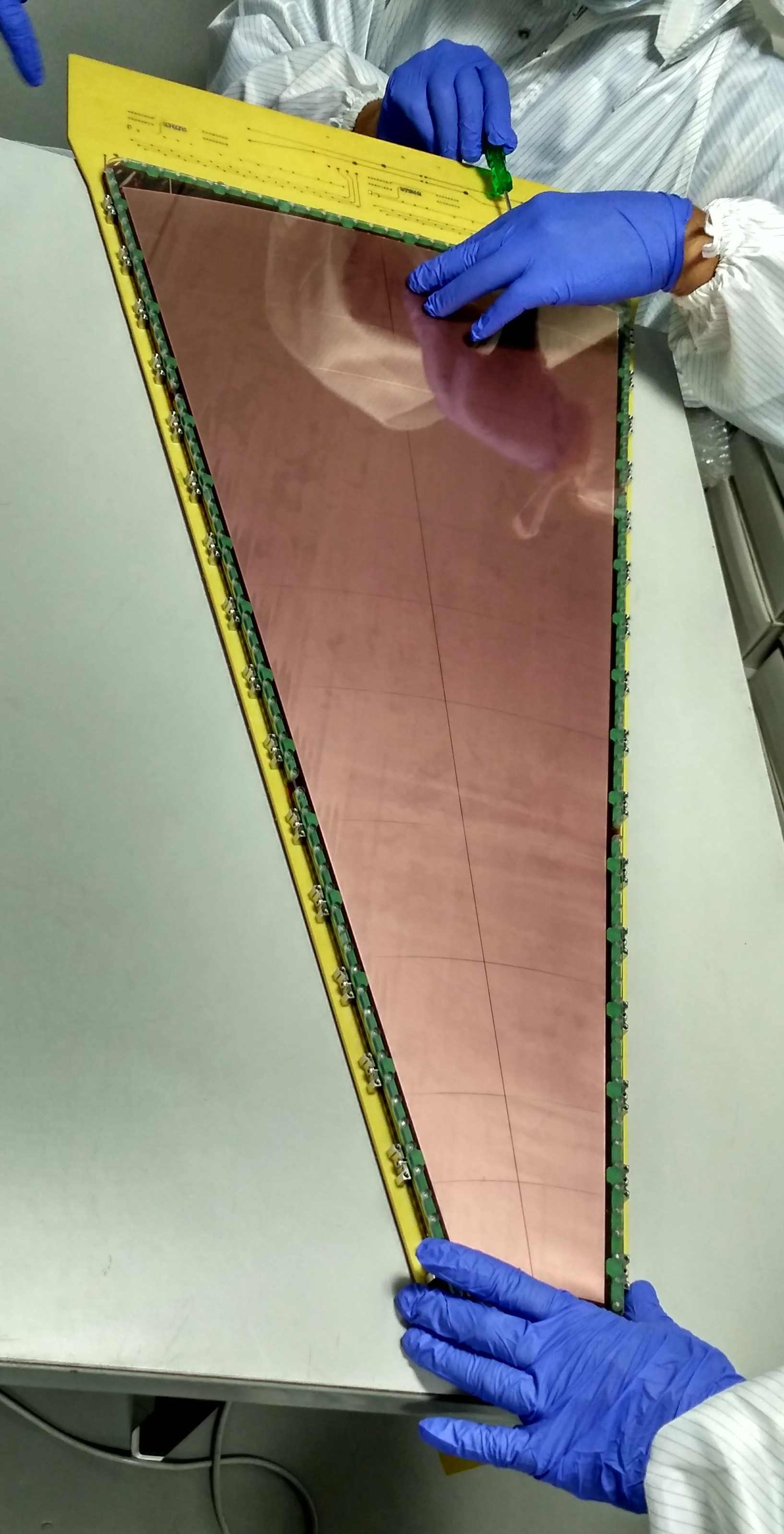}
\hspace{3mm}
\includegraphics[width=3.5cm,height=5.5cm]{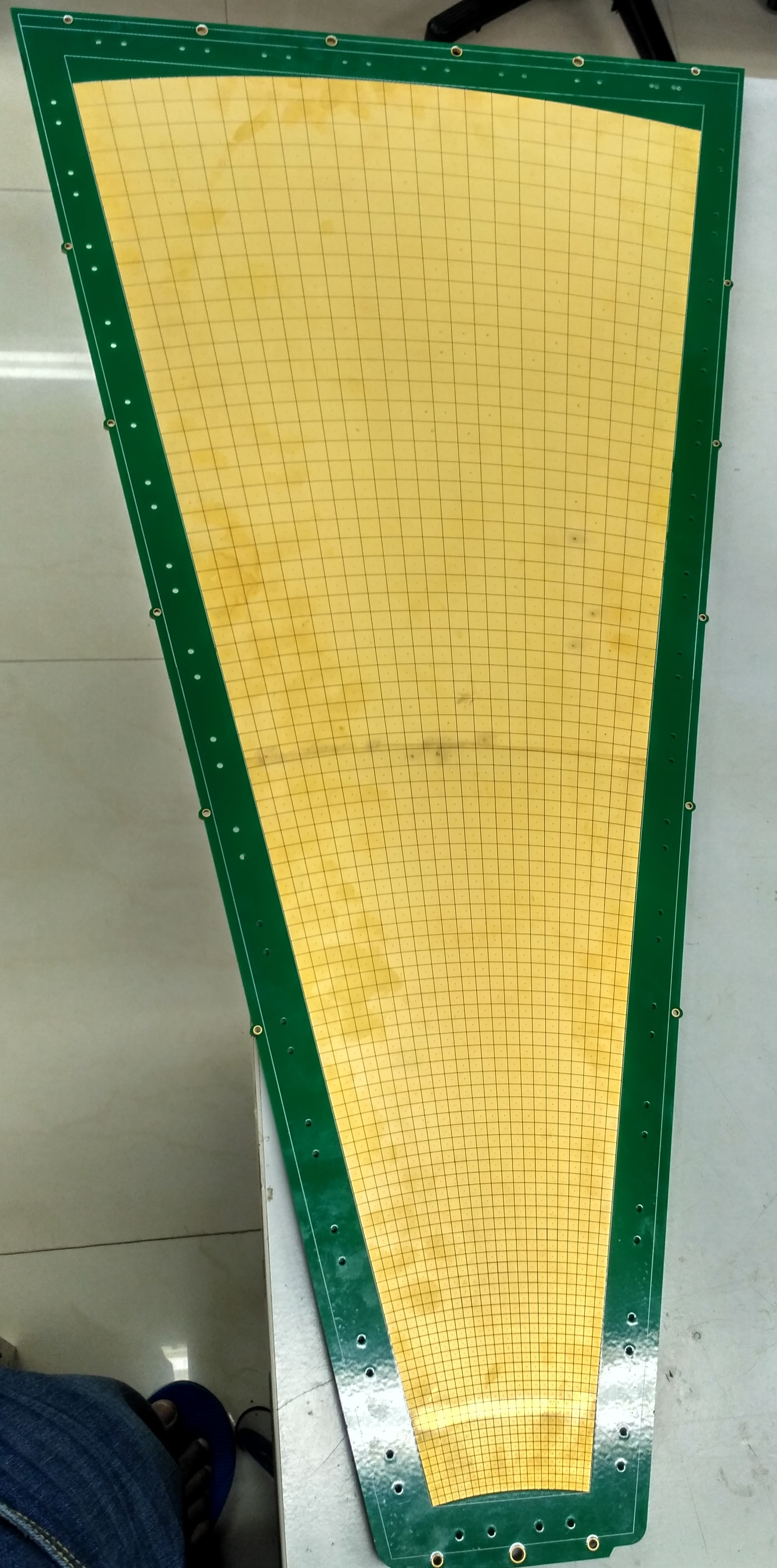}
\vspace{-2mm}
\caption{ Left: Schematic of MuCh system in CBM. Middle: Chamber assembly. 
Right: Picture of the trapezoidal readout board for 1$^{st}$ station of MuCh.} \label{fig:MuchSchematicAndreadoutPCB}
\end{figure}

The Muon Chamber (MuCh) system~\cite{MuChTDR:161297} will be used to measure the dimuon signals originating from the heavy-ion collisions, which is one of the diagnostic probes to understand the physics of the fireball. MuCh consists of a series of segmented absorbers and detector stations sandwiched between them. A triplet of detector layers will be used in each station. A schematic diagram of the MuCh setup is shown in the left of Fig.~\ref{fig:MuchSchematicAndreadoutPCB}, showing the detector layers and segmented absorbers. In the first two stations, owing to high particle flux, high rate detectors based on Gas Electron Multiplier (GEM) technology will be used~\cite{DUBEY2013418, DUBEY201462, ADAK201729}. GEM detectors have been chosen for particle detection in many experiments ~\cite{CERN-LHCC-2017-012, CERN-LHCC-2013-020, Cardini:1495070, Klest_2020, ANDERSON201135}. Likewise, in the CBM experiment, the role of the large area triple GEM detectors of MuCh is to carry out charged particle tracking for muon identification. The specifications of the GEM modules for the first two stations of MuCh are given in table~\ref{tab:table}. In any given layer, since the particle densities per unit area change when going radially outwards in the transverse direction from beam pipe, the pad-granularity of the GEM-MuCh module monotonically decreases from $\sim$3.2~mm pads in the inner region to about $\sim$17~mm on the periphery of the trapezoid for the modules of the first station, as can be visualized in Fig.~\ref{fig:MuchSchematicAndreadoutPCB} (right).

\begin{table}[htbp]
\label{tab:table}
\centering
\caption{\label{tab:1} Table for the specification of modules of the first two stations.}
\begin{tabular}{|c|c|c|c|c|c|c|}
\hline
Station   & Number       & Total \#      &  Total \#              & Total \#      & Typical dimension    & Granularity\\
number    & of           &  of     &   of readout         &  of    & of one module (cm)   & (mm)\\
		  & layers		  & modules	&   channels	  &  FEBs    & (active area)        & \\
\hline
1  & 3  & 48 & $\sim$107k & 864 & Inner $\sim$7.5    & Min. $\sim$3.2  \\
   &    &    &            &     & Outer $\sim$40.0   & Max. $\sim$17 \\
   &    &    &            &     & Length $\sim$80.0  &  \\
\hline
2  & 3  & 60 & $\sim$109k & 900 & Inner $\sim$7.9    & Min. $\sim$4.5 \\
   &    &    &            &     & Outer $\sim$41.5   & Max. $\sim$21.5 \\
   &    &    &            &     & Length $\sim$100.5 &   \\
\hline
\end{tabular}
\end{table}

All detector systems of CBM will have to cope with very high interaction rates up to 10\,MHz, in order to detect rare probes~\cite{CBMAblyazimov2017}. The maximum expected particle rate per unit area for the first station of MuCh is about 400~kHz/cm$^{2}$ for minimum bias Au+Au collision at 10~AGeV beam energy. The signals from all the CBM detector subsystems will be read out using self-triggered electronics, enabling one to collect data at this high rate, whereby every signal above a set threshold is recorded along with the timestamp of these hits. Events are then reconstructed offline by grouping such hits in time. It, therefore, becomes essential to study an integrated response of the detectors through a series of beam tests in which one studies in detail the performance of all the detector subsystems of CBM prior to the actual experiment. In this regard, a precursor experiment consisting of all the CBM detector subsystems has been setup at the SIS18 facility of GSI called mCBM~\cite{mCBMExp} (“mini-CBM”) as a part of the FAIR Phase 0 program. The mCBM experiment aims as a CBM full-system test and enables tests of real-size modules of each detector subsystem under realistic experimental conditions in high-rate nucleus-nucleus collisions. Major tasks are studying and optimizing the DAQ and data transport to a computer farm, in particular, the timing-stability and data consistency of the subsystem data streams, investigating issues related to the operation of detectors in a high-rate environment, and developing and optimizing the software for online/offline data analysis. In addition to this, the reconstruction of rare events like $\Lambda$ Hyperon production in nucleus-nucleus collisions at SIS18 energies will be performed.

In view of this, two real-size triple GEM detectors corresponding to the module sizes of station-1 of MuCh were installed and commissioned in the mCBM experiment. Ar/CO$_{2}$ (70/30) gas mixture has been the fill gas for the present tests. Tests with real-size detectors have been reported in~\cite{ADAK201729, Kumar:2017noz}. However, the earlier tests were carried out with single-particle proton beams focused at a specific location on the detector. At mCBM, the goal is to study the simultaneous response from different regions of the detector and in conjunction with those from other subsystems when a multitude of particles originating from nucleus+nucleus collisions pass through them. The first such attempt was carried out at CERN-SPS~\cite{Kumar:2017noz}, where the real-size modules used were of different gap-configuration and were operated with different electronics compared to the present modules used in mCBM. Moreover, the data at mCBM have been taken using the upgraded CBM DAQ~\cite{mCBMProposal} and final readout ASIC, STS/MuCh{-}XYTER electronics~\cite{KASINSKI2018225, Kleczek_2017}, which has been specifically designed and developed for use by STS/MuCh subsystems in CBM.

Data were taken with Ar beams colliding on Au target of thickness 2.5~mm in November and December 2019 beamtime. Preliminary results from the GEM-MuCh modules at mCBM have been reported in~\cite{Kumar_2020}. In this paper, we report test results with the latest version of  STS/MuCh{-}XYTER, i.e., v2.1. The results discussed in this paper correspond mainly to a detector gain of about 3.1 x 10$^{3}$. The details of the GEM detector modules, their design, and fabrication are given in section~\ref{sec:DesignOfPrototype}.  The schematic of the experimental setup is described in section~\ref{sec:ExpSetup}. The basic response of the GEM chambers in terms of observing the spill structure using GEM modules, studying the time synchronization of the detector hits, and the study of detector characteristics such as detector gain, cluster size, etc., have been reported in section~\ref{sec:AnalysisAndResults}. The factors affecting the timing characteristics have also been discussed in this section. We finally summarize in section~\ref{sec:Summary}.



\begin{figure}[htbp]
\centering 
\includegraphics[width=7.0cm,height=7.0cm]{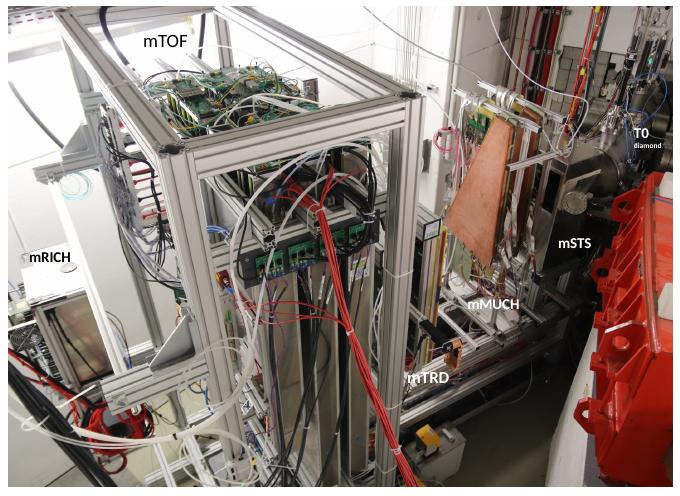}
\hspace{5mm}
\includegraphics[width=4.0cm,height=7.0cm]{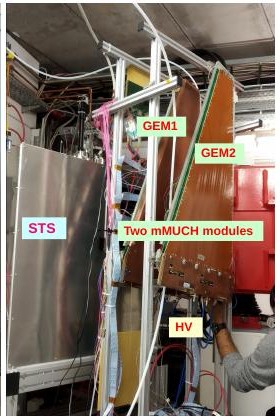}
\vspace{-3mm}
\caption{\label{fig:mCBMSetupPicture} Left: Photograph of the experimental setup as of November 2019 
showing the installed detector subsystems inside the mCBM cave at the SIS18 facility of GSI. 
Right: Closeup picture of two installed mMuCh modules.}
\end{figure}

\section{Design and fabrication of prototype GEM detectors}
\label{sec:DesignOfPrototype}
Each of the two sector-shaped GEM chambers of mCBM consists of three GEM foils stacked in a 3-2-2-2 gap configuration, where the number represents the gap between different layers in mm. These modules have been designed and fabricated at VECC, Kolkata. Large size trapezoidal-shaped, single-mask GEM foils were procured from CERN. These were stretched using the ``NS-2'' technique~\cite{NS2Technique}, which does not use glue. The top surface of each foil is segmented into 24 divisions. As part of essential Quality Assurance (QA) of the GEM foils, the foils were selected only when they satisfied the criteria of no-short along with a low leakage current of <5 nA at $\Delta$V$_{GEM}$ of 550~V for every segment. The measurement was done in air at an ambient temperature of about 23$^\circ$C and RH of around 45-50$\%$. A novel optocoupler-based HV biasing scheme coupled to two resistive chains was adopted for every module to power all the 24 foil segments. The optocoupler aims to isolate any segment that may suddenly go bad in the course of operation, thus enabling the module to function with the rest of the segments. Currently, ``Dip-switches'' are provided on the extended portion of the drift PCB to isolate any segment whenever required physically. These, in the future, would be coupled to micro-controller-based devices to control the on-off operation remotely. In the mCBM campaign so far, no segment short has been observed in the GEM-MuCh modules. A detailed discussion on optocoupler based design can be found in~\cite{GEMAjitOpto2019}. More details of the steps of fabrication have been discussed in~\cite{ADAK201729}. The active area of each of the trapezoidal detectors is about 1900 cm$^{2}$, corresponding to 2231 pads. The entire detector module is read out using 18 front-end boards (FEBs), each of which was connected to the pads in different regions on the detector plane.     

\section{Experimental setup}
\label{sec:ExpSetup}

\begin{figure}[htbp]
\centering 
\includegraphics[width=8.0cm,height=6.0cm]{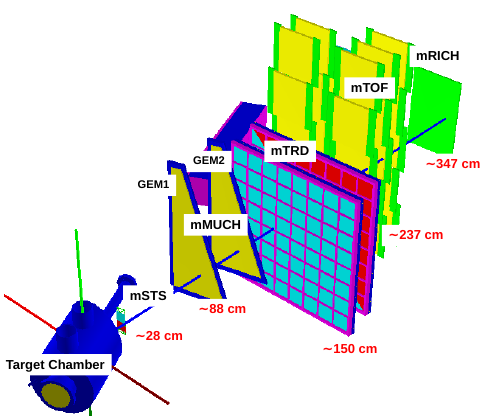}
\vspace{-4mm}
\caption{\label{fig:SchematicSetupmCBM} Schematic setup of mCBM experiment as of November 2019. 
Detectors are placed at approximately 25$^{\circ}$ from the beam axis. The diamond detector (T0) is placed inside the target chamber. 
mSTS - mini-Silicon Tracking System, mMuCh - mini-Muon Chamber System, mTRD - mini-Transition Radiation Detector, 
mTOF - mini-Time Of Flight, mRICH - mini-Ring Imaging Cherenkov.}
\end{figure}

A photograph of the two GEM modules as of November 2019, installed along with other detector subsystems inside the mCBM cave located at the SIS18 facility of GSI, is shown in Fig.~\ref{fig:mCBMSetupPicture} (left). The subsystems were placed along the detector axis (25$^{\circ}$ line from beam direction) at different distances from the target, the position of which is taken as the origin (0,0,0). All the detector centers were (approximately) positioned at the height of 2\,m from the ground and 25$^{\circ}$ from the beam axis. Both mMuCh modules were mounted on an Al-plate of $\sim$12\,mm thickness. Each plate consisted of $\sim$6\,mm Al-pipe winding inside the plate for carrying chilled water, thus providing the cooling of the front-end boards (FEBs)~\cite{MuchCooling}. A total of 18 FEB's (front-end-boards) are needed to populate one full chamber of station~1, with each FEB dissipating 2.5\,W of heat. However, for the data reported in this paper, the number of working FEBs was around 16 and 9 in GEM1 and GEM2, respectively. For the ease of installation, the dimensions of the Al-plates used in mCBM are slightly different from the actual design for the CBM experiment. The entire mMuCh system was then installed on the beam table, as shown in Fig.~\ref{fig:mCBMSetupPicture} (right). A schematic of the mCBM setup is shown in Fig.~\ref{fig:SchematicSetupmCBM}, typical distances of the detector subsystems are as shown. The diamond (T0) detector is placed inside the target chamber. The dimension of the T0 detector is approximately 20~mm~x~20~mm and placed roughly at 20~cm upstream to the target. The mMuCh detectors were positioned with the readout side facing the target and oriented such that the long trapezoidal side was along the vertical axis, as shown in the figures. The detector axis intersected the modules at a distance of about 20\,cm from the bottom edge of the trapezoid.

Signals from the readout pads of the mMuCh modules were sent via the readout connectors to the corresponding FEBs fixed on the Al-cooling plate, using 10\,cm long flexible Kapton cables. These FEBs were then connected to the CBM DAQ system~\cite{mCBMProposal, Kumar_2020}. In a free-streaming mode, signals from every electronic channel, whenever higher than a set threshold, are recorded along with their timestamps. The digitized signal from any electronic channel is referred to as Digi. The readout ASIC has 128 analog channels. Each channel uses a 5-bit Flash ADC, which was calibrated with a step size of 2.5\,fC. Suitable thresholds were set to control the noise. Few channels in both modules were found to have exceptionally high noise and were masked before starting the data run. The ASIC does have a provision to set different thresholds to specific channels. However, implementing this in situ involves unplugging FEBs and re-calibrating the concerned channels, which is cumbersome. For studying the response from different areas of the detector at the same footing, a configuration with a common threshold for all channels is always preferred. Based on the observed noise behavior of the chambers, the thresholds of all the FEE boards were set to 6\,fC, such that the noise rate stayed within the acceptable range of the usable bandwidth of the FEE boards. However, for a few boards in GEM2, a higher threshold had to be chosen.

\section{Data analysis and first results}
\label{sec:AnalysisAndResults}

\begin{figure}[htbp]
\centering 
\includegraphics[width=9.0cm,height=5.5cm]{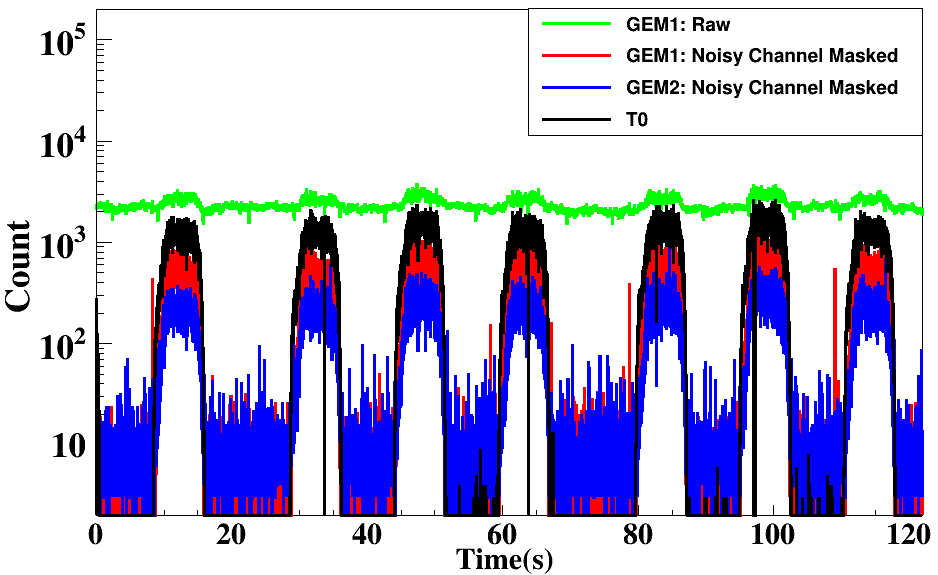}
\includegraphics[width=6.0cm,height=5.5cm]{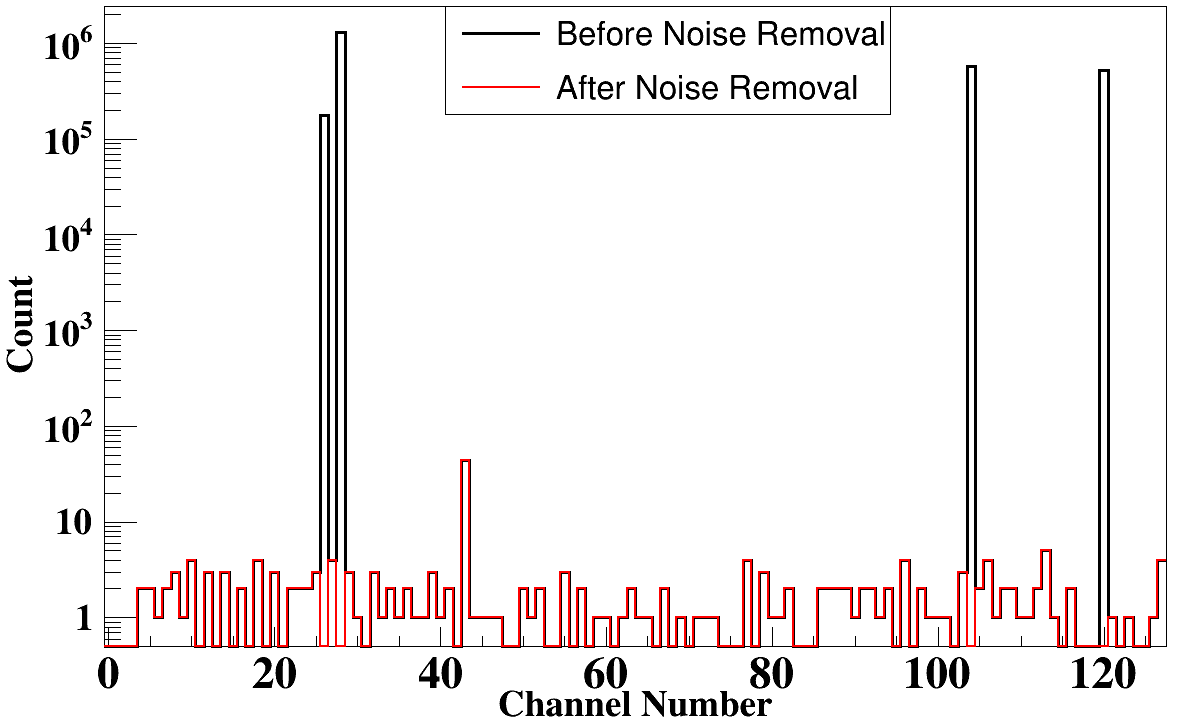}
\vspace{-8mm}
\caption{\label{fig:SpillStructureAndChannelDist} Left: Variation of Digi counts with time, spill structure, for GEM1-raw (green), 
GEM1/GEM2 - noise subtracted and T0. Right: Channel hit distribution for one of the FEB's in beam-off condition. Black corresponds 
to the raw channel count, and red is after masking noisy channels. Only a few channels are found to be noisy.}
\end{figure}

We report here the first results from data taken in November and December 2019 with $^{40}$Ar beam colliding on an Au-target of 2.5\,mm thickness at an average intensity of 5$\times$10$^{6}$ per spill. A preliminary CBM data analysis chain based on the CbmRoot software framework \cite{CbmRootFram} has been applied to the data. This involves raw data unpacking, including individual time-offset corrections for the detector subsystems (digis), a first event building by a timestamp cluster search and hit reconstruction for the detector subsystems. Using the Digi as well as the reconstructed hits, a dedicated analysis procedure was used to extract the mMuCh performance in dependence of various detector and analysis parameters.

The detector was operated at a summed GEM voltage of about 1072\,V. The data were grouped in a time slice of size 10.24\,ms. The typical distribution of signal counts from the GEM1 detector in time-bins (10\,ms), taking the first hit time as a zero reference, is shown in the left panel of Fig.~\ref{fig:SpillStructureAndChannelDist}. The incident beam is in the form of spills, as displayed by the distribution of T0 counts. One notices that the raw counts in GEM1 (in green) do not display a proper spill structure. This is because of the large noise contribution, which is dominant in both on-spill as well as off-spill regions. An offline investigation of the off-spill data revealed that a few channels in some of the FEBs produced large noise, as can be seen from the channel-wise counts distribution for one FEB of GEM1 in Fig.~\ref{fig:SpillStructureAndChannelDist} (right). With the help of off-spill data, such noisy channels were identified in all ASIC's and were accordingly masked for further analysis. The signal-counts distribution, after removal of noisy channels for GEM1 (red), GEM2 (blue) reveal the spill structure (Fig.~\ref{fig:SpillStructureAndChannelDist} (left)). They are now observed to be well correlated with those of T0, as expected. The residual count still remaining in the off-spill region is indicative of the noise rate and is observed to vary from one FEB-to-FEB. Also, a higher noise level for GEM2 was observed as compared to GEM1. However, these are of much lower counts than the counts in the on-spill region.   

\begin{figure}[htbp]
  \centering
  \includegraphics[width=4.5cm,height=4.2cm]{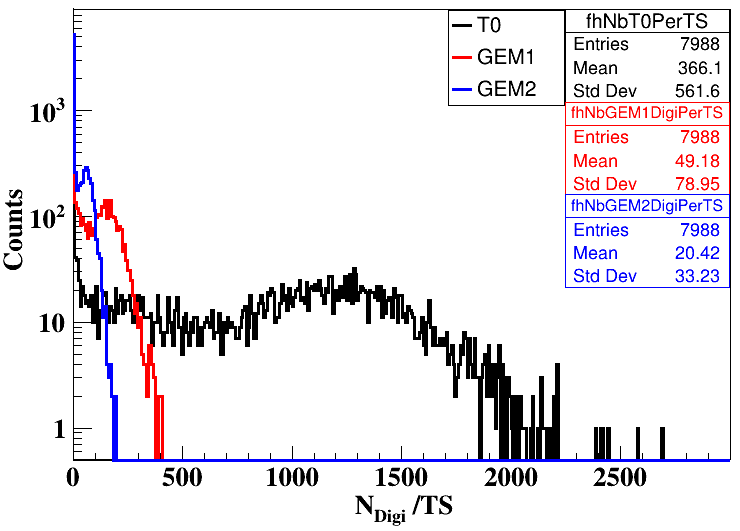}
  \includegraphics[width=4.5cm,height=4.2cm]{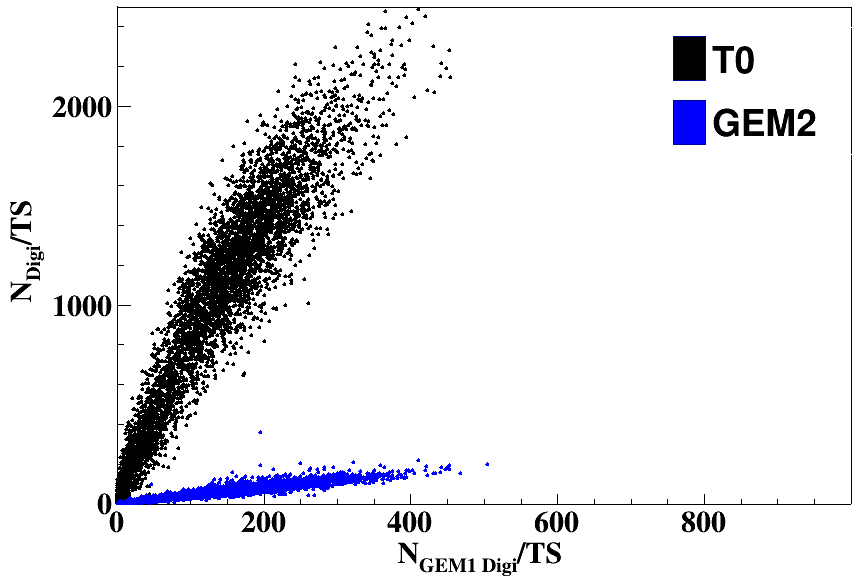}
  \includegraphics[width=4.5cm,height=4.2cm]{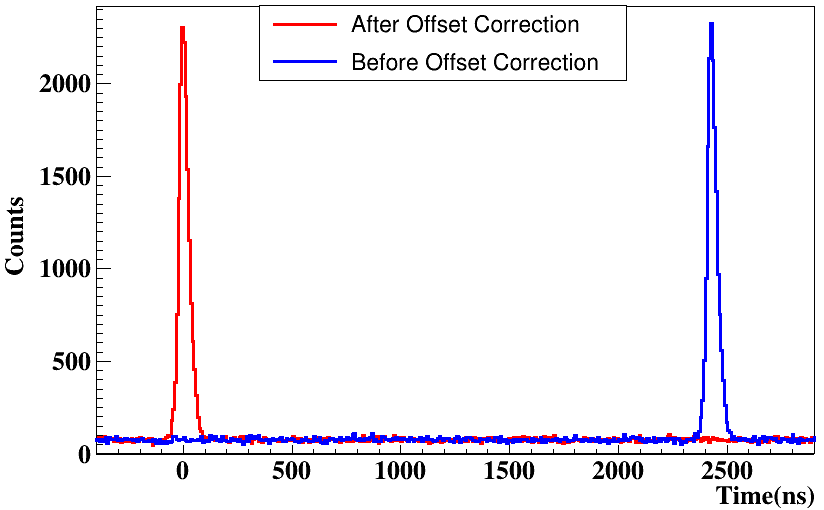}
  \vspace{-3mm}
\caption{\label{fig:DigiCorrInTS} Left: Distribution of the number of digis per time-slice (Digi/TS) for GEM1 (red), GEM2 (blue) and T0 (black). 
Middle: Variation of the number of T0 Digi/TS with GEM1 Digi/TS (red) and the number of GEM2 Digi/TS with GEM1 Digi (blue). 
Right: Time difference distribution between one FEB of GEM1 with T0 before (blue) and after (red) offset correction (see text).}
\end{figure}


The extracted spill length of the beam is about 7\,s, with off-spill lengths of approximately 7\,s and 12\,s, (parallel user operation) as can be noticed in Fig.~\ref{fig:SpillStructureAndChannelDist}, left side. As has been mentioned, the data were collected in 10\,ms Time-Slice (TS) intervals. The distribution of the number of digis per TS for the two GEM modules and T0 is shown in Fig.~\ref{fig:DigiCorrInTS} (left). This includes counts in the off-spill region as well, which being relatively much lower, show up in the form of a peak at values around 10-30 Digi/TS for both the GEM modules, while the peaks on the higher side of the spectra are due to signals from on-spill. A difference between these two peaks provides the average number of digis/TS for any detector, corresponding to the particular operating conditions at which the data was taken. The on-spill peak for GEM2 is lower than that of GEM1 because of its lower acceptance owing to the smaller number of working FEBs. The correlation in terms of the number of digis per TS between the three subsystems is shown in the middle panel of the same figure.

Each collision is a point in time. Hence the digis from different detector subsystems should be correlated in time. This is revealed from the time difference spectra shown in Fig.~\ref{fig:DigiCorrInTS} (right, blue) where, within every time slice interval, the time difference distribution between the digis from a FEB of GEM1 and T0 has been plotted. This represents a typical time-correlation spectra. The mean position of this distribution (around 2500\,ns) represents the offset for the particular FEB, and this varies from one FEB to another. Synchronizing all the hits from different FEBs becomes thus essential for a proper data analysis. In this direction, time offset for all the FEBs corrections need to be properly determined. The offset-value is subtracted from timestamps of each Digi for the corresponding FEB, such that the resulting time-corrected distribution now peaks at zero, as per construction (represented by the red curve in Fig.~\ref{fig:DigiCorrInTS} (right)). Such offset corrections were applied for all subsystems separately.


\begin{figure}[htbp]
\centering 
\includegraphics[width=4.5cm,height=4.0cm]{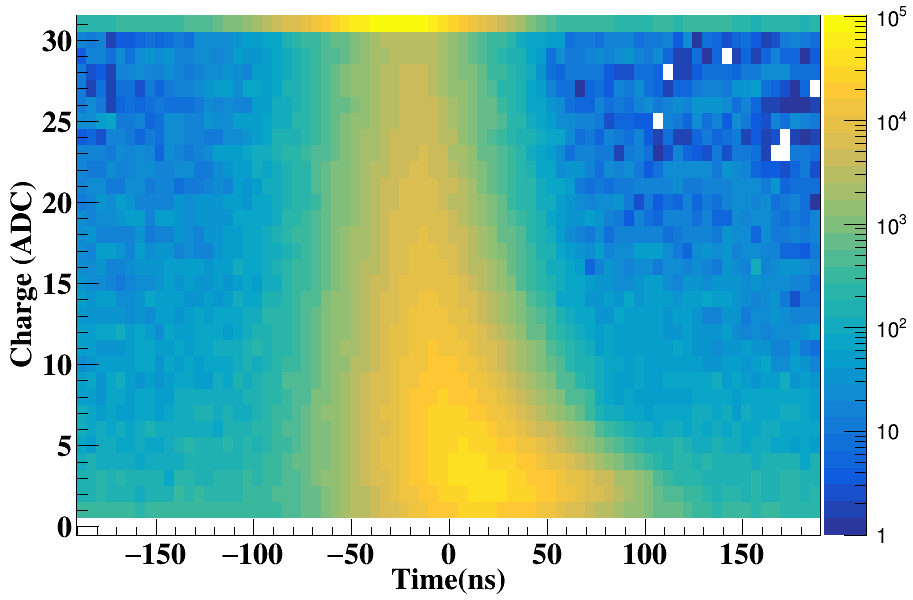}
\includegraphics[width=4.5cm,height=4.0cm]{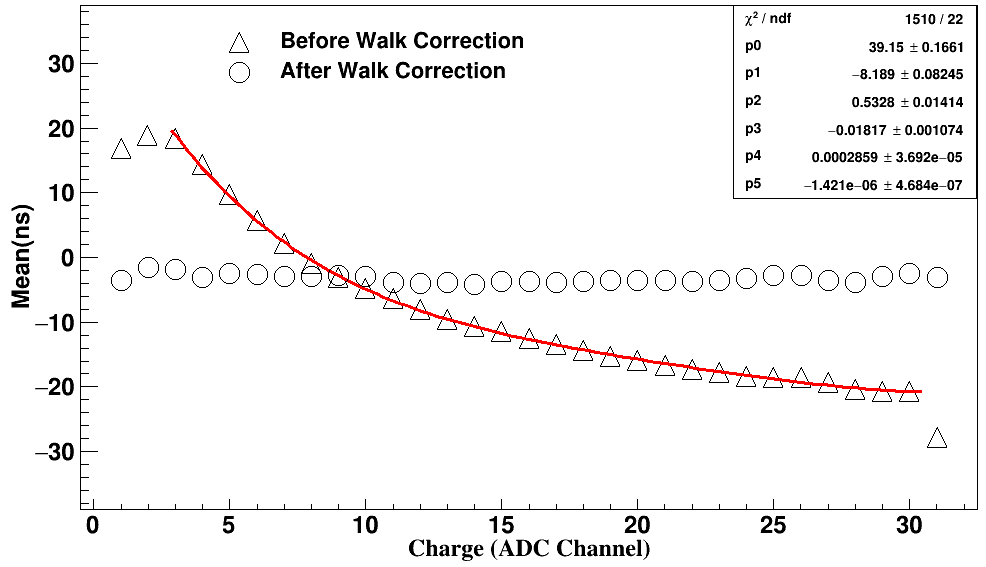}
\includegraphics[width=4.5cm,height=4.0cm]{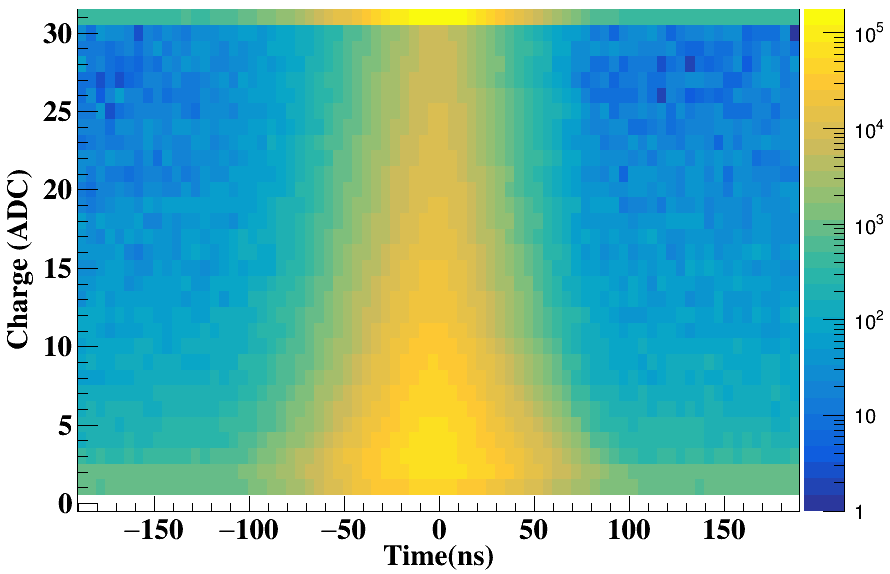}
\vspace{-3mm}
\caption{\label{fig:WalkCorrection} Left: Variation of the time difference between signals in GEM1 and one of the TOF modules (super module 0) with pulse height (in ADC unit) of GEM1.  
Middle: Variation of the mean (ns) of the time difference spectra with ADC before (upper triangle) 
and after (open circle) time-walk correction. The raw values are parameterized using a polynomial fit. 
Right: Variation of the time difference between GEM1 and one of TOF counter for different ADC values after time-walk correction.}
\end{figure}

\begin{figure}[htbp]
\centering 
\includegraphics[width=6.5cm,height=4.5cm]{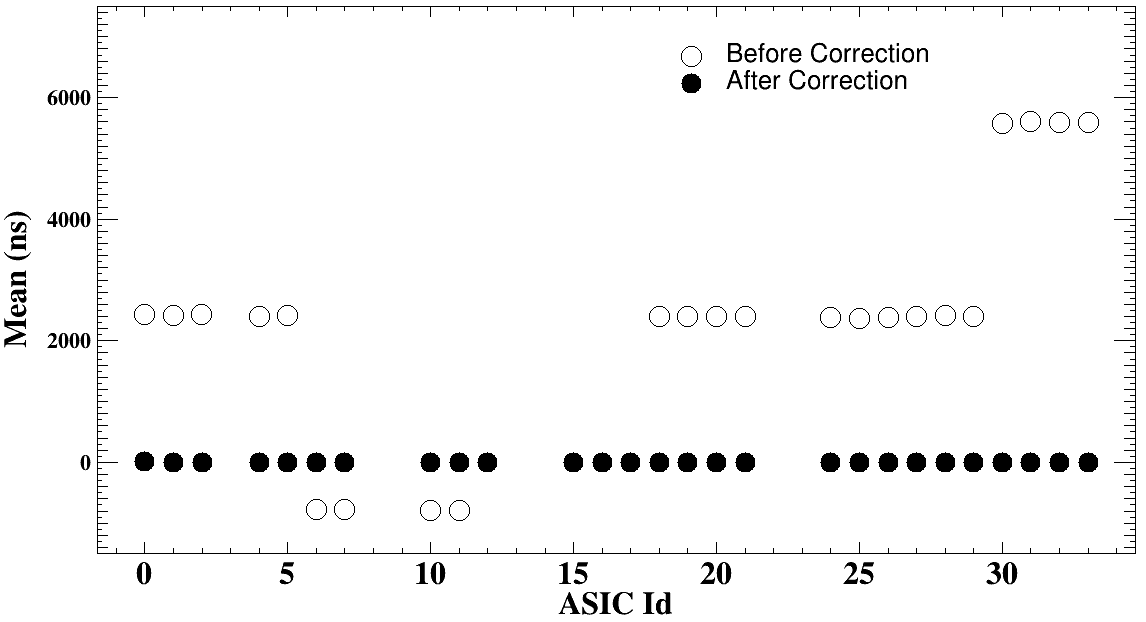}
\includegraphics[width=6.5cm,height=4.5cm]{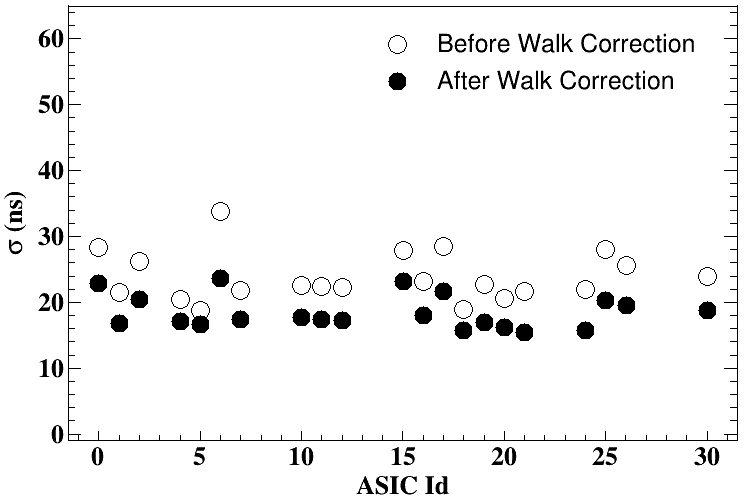}
\vspace{-3mm}
\caption{\label{fig:MeanAndSigmaVariationAllFEBs} Left: Variation of the mean (ns) of time difference spectra for each ASIC with ASIC number 
before (open circle) and after (close circle) offset correction. Mean positions for a few ASICs are not shown here 
due to large offset value (several $\mu$s).  Right: The variation of $\sigma$ with ASIC number 
before (open circle) and after (close circle) time-walk correction. The ASIC Id for GEM1 is from 0 to 23, while those beyond belong to GEM2.}
\end{figure}

The time correlation spectra are obtained for each FEB separately, and the extracted peak-position after appropriate Gaussian fit provides the time-offset for all the channels of the corresponding FEB. By implementing these FEB-to-FEB corrections on the entire data set, we have made an effort to bring all the digis synchronization with those from T0 for further analysis. However, there could still be time-walk effects~\cite{KASINSKI2018225} on the signal timing due to varying signal strengths, as can be seen from the 2D distribution of ADC vs. time difference for GEM1 in Fig.~\ref{fig:WalkCorrection} (left). The variation of the time difference peak position (ns) for the corresponding ADC bin is shown in Fig.~\ref{fig:WalkCorrection} (middle). It has been fitted with a polynomial to find the parametric equation for the time-walk correction. The 2D distribution after time-walk correction is shown in Fig.~\ref{fig:WalkCorrection} (right). The mean offset values for every FEB is now well peaked at around zero ns, by construction, as depicted in Fig.~\ref{fig:MeanAndSigmaVariationAllFEBs} (left). Open circles denote the time-offset values for different FEBs (represented by ASIC Id on the X-axis). For a few FEBs, these offsets were very high (in several $\mu$s) and exceeded the scale of the plot. Eventually, all these offsets could be corrected (full circles). The width of the time difference spectra provides the time resolution of the detector. Using a Gaussian fit, the sigma ($\sigma$) of this distribution for both before and after the time-walk correction has been shown in Fig.~\ref{fig:MeanAndSigmaVariationAllFEBs} (right). An improvement of 6-7\,ns is observed after correction.


\begin{figure}[htbp]
\centering 
\includegraphics[width=7.0cm,height=6.0cm]{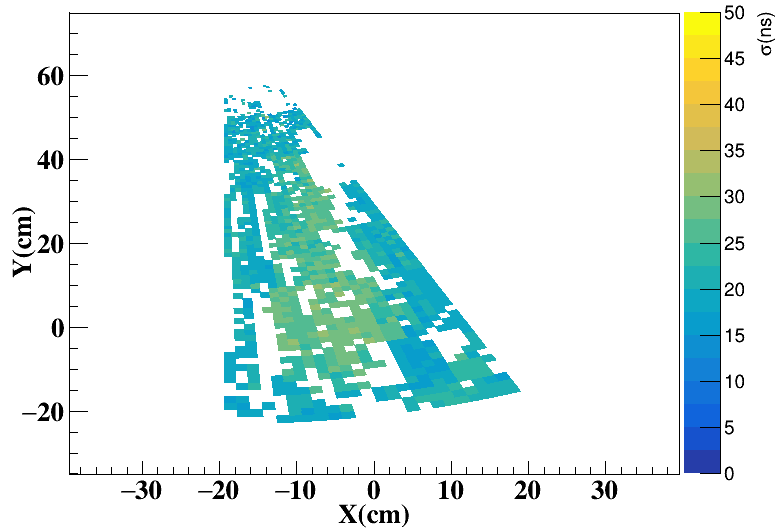}
\hspace{2mm}
\includegraphics[width=7.0cm,height=6.0cm]{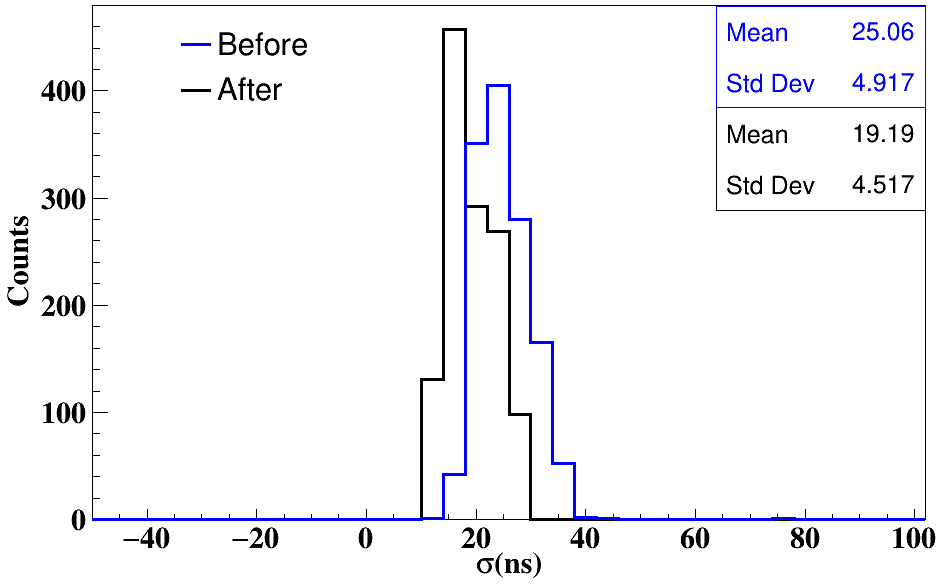}
\vspace{-3mm}
\caption{\label{fig:MeanAndSigmaDistributionOfAllChannelsOfGEM1} Left: Time resolution map on GEM1 plane. The z-axis is the $\sigma$ (ns) of each pad. 
Right: 1-D distribution of time resolution for all the pads (shown left) before (Blue) and after (Black) time-walk correction.}
\end{figure}

A deeper look at detector uniformity has been carried out at the level of pads/channels. The 2D-plot graphically display the vatiation in time resolutions (after time-walk correction) for about 1400 channels of GEM1 in Fig.~\ref{fig:MeanAndSigmaDistributionOfAllChannelsOfGEM1} (left). This variation is observed to be of the order of 4-5~ns from the mean value, which can be seen from the width of the 1-D distribution of $\sigma$ (right panel of Fig.~\ref{fig:MeanAndSigmaDistributionOfAllChannelsOfGEM1}).

In a self-triggered system, event building, as well as reconstruction, is a challenging task. Using the timestamps of the signals from the detectors, an algorithm of building events by suitably grouping digis in time was used.  As a first attempt in this direction, we have used a time window of 200\,ns to group together the digis along with a condition of having at least one~T0 and six TOF digis within this time interval. This time window of 200\,ns was considered based on the typical time resolution of the detector subsystems as well as on rather moderate collision rates. All MuCh digis inside this time window are clubbed together with the T0 and TOF digis forming events subsequently.  

\begin{figure}[htbp]
\centering 
\includegraphics[width=4.0cm,height=4.0cm]{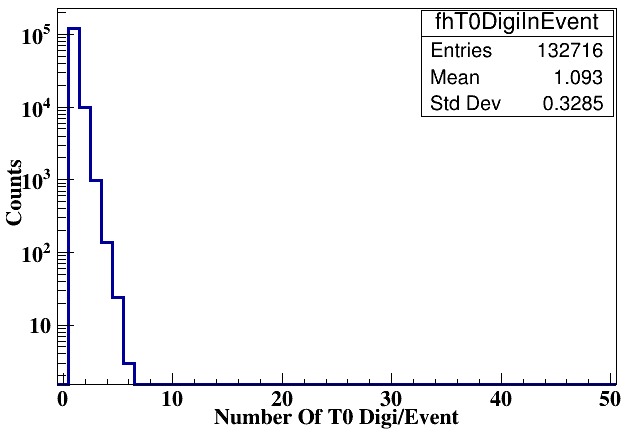}
\includegraphics[width=4.0cm,height=4.0cm]{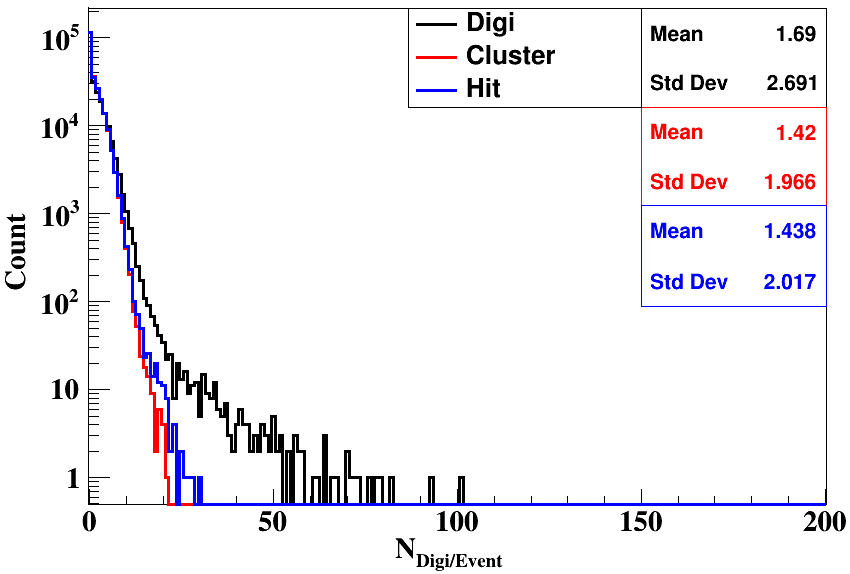}
\includegraphics[width=4.0cm,height=4.0cm]{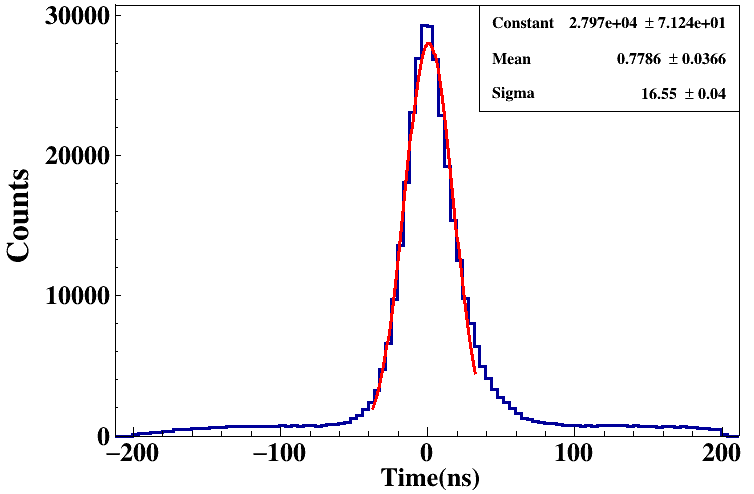}
\vspace{-2mm}
\caption{\label{fig:AvgT0DigiAndMuchDigiCluHits} Left: Distribution of the number of T0 Digi / event. 
Middle: Distribution of number digi (black), Cluster (red) and Hits (blue) per event. 
Right: Time difference distribution between hits of GEM1 with T0 digi in the event.}
\end{figure}

As mentioned, the event building algorithm is still under investigation. One quality check for the algorithm is to study the T0 distribution event by event. Ideally, for every collision, there should be one T0 Digi. Each time window satisfying the hit-criteria is supposed to represent one collision, implying that the number of T0 digis per event should typically average around ``1''. This is revealed in Fig.~\ref{fig:AvgT0DigiAndMuchDigiCluHits} (left), where the average number of T0 digis per event yields to be about 1.1 (matching with our expectation within 10~\%).

The event building was followed by cluster finding and hit reconstruction within the MuCh modules. Each particle track that produces a signal in the detector may affect one or more pads (digis). A nearest neighbor algorithm was used to club together with the digis in an event forming a fired-pad clusters. Clusters having more than one local maximum were split further, thus forming ``Hits''. The middle panel of Fig.~\ref{fig:AvgT0DigiAndMuchDigiCluHits} shows systematically the event-by-event distributions of the number of digis, the number of clusters, and finally, the number of ``Hits'' for GEM1. As a result of the cluster splitting, the distribution of ``Hits'' is observed to be marginally wider than that for the clusters. The time correlation of the MuCh ``Hits'' with T0 digis within an event is shown in Fig.~\ref{fig:AvgT0DigiAndMuchDigiCluHits} (right). During ``Hit'' reconstruction, the algorithm chooses the smallest time of the Digi within the cluster; hence, the width of this time-correlation spectra is slightly smaller than that in Fig.~\ref{fig:MeanAndSigmaDistributionOfAllChannelsOfGEM1} (right). The measured time resolution values at mCBM are pretty close to what has been measured using single-particle beams \cite{ADAK201729}. The width of the time correlation spectra would play a crucial role in optimizing the time window of the event building algorithm, particularly when operating at high collision rates.

\begin{figure}[htbp]
\centering 
\includegraphics[width=6.5cm,height=5.8cm]{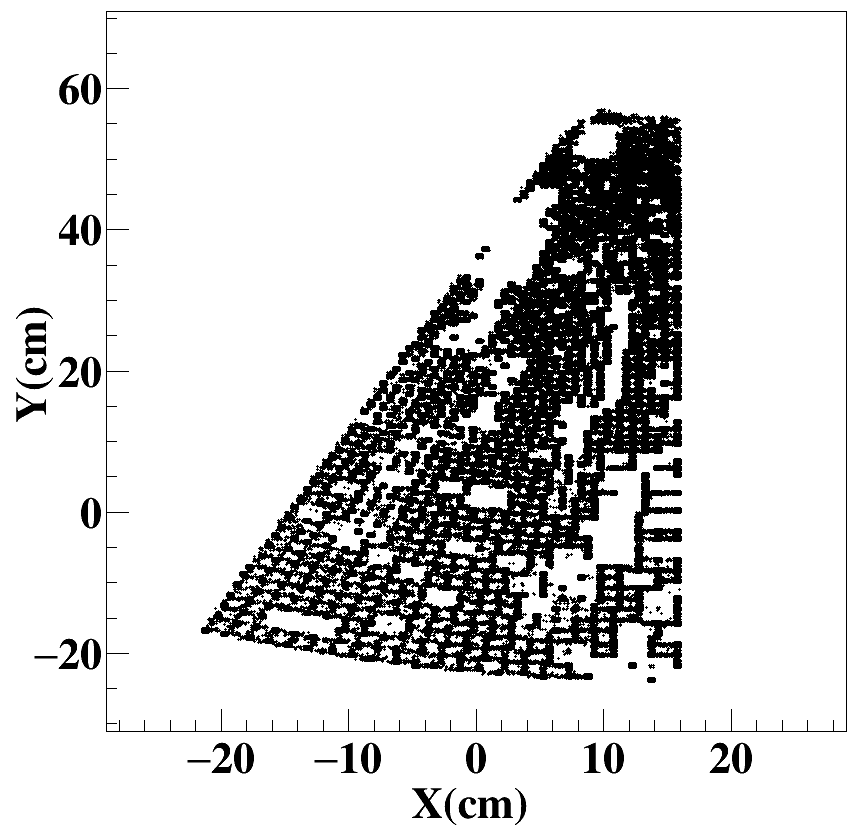}
\includegraphics[width=6.5cm,height=5.8cm]{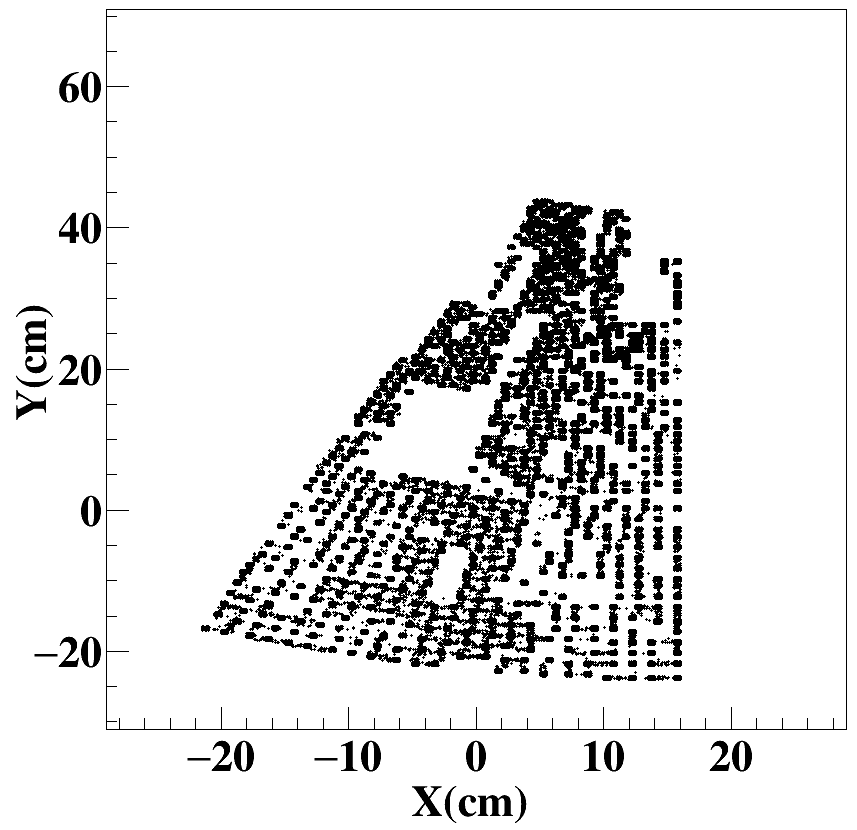}
\vspace{-3mm}
\caption{\label{fig:XYDistHitsG1G2} X (cm)-Y (cm) distribution of hits for GEM1 (left) and GEM2 (right) after hit reconstruction. The white spaces mostly correspond to the dead areas, mainly due to non-working FEBs and masked noisy pads/channels in the detector.}
\end{figure}

\begin{figure}[htbp]
  \centering
  \includegraphics[width=4.5cm,height=4.0cm]{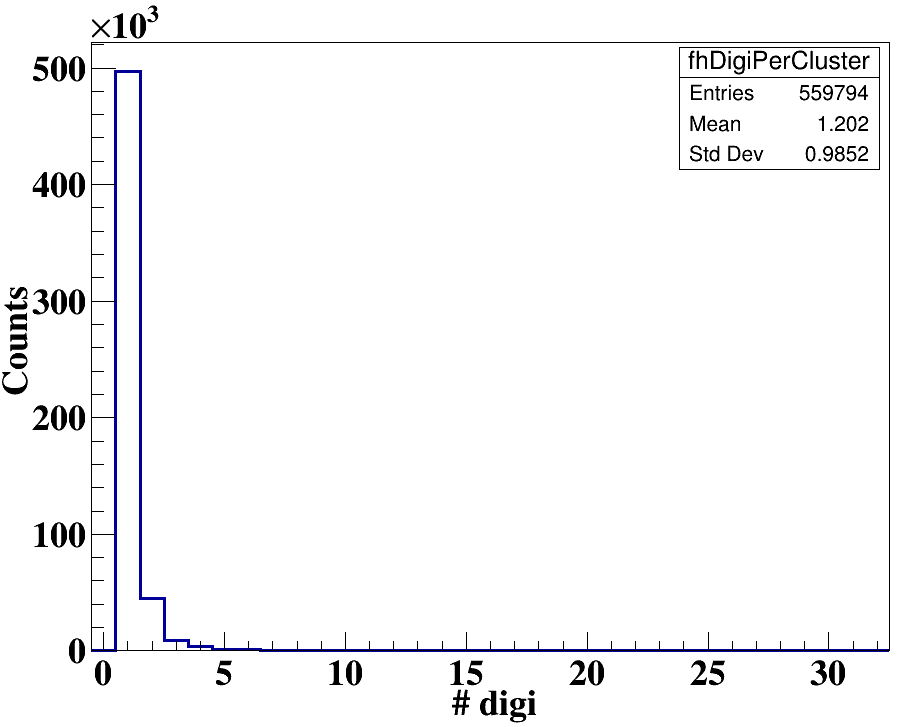}
  \includegraphics[width=4.5cm,height=4.0cm]{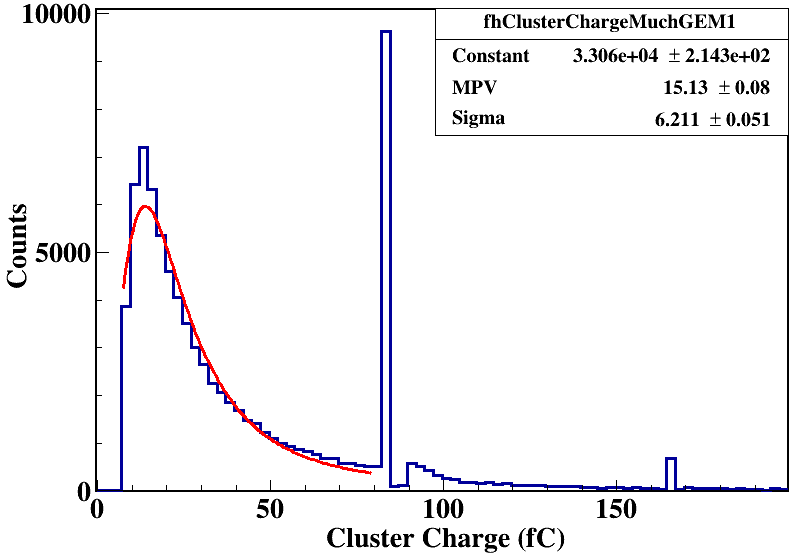}
  \includegraphics[width=4.5cm,height=4.0cm]{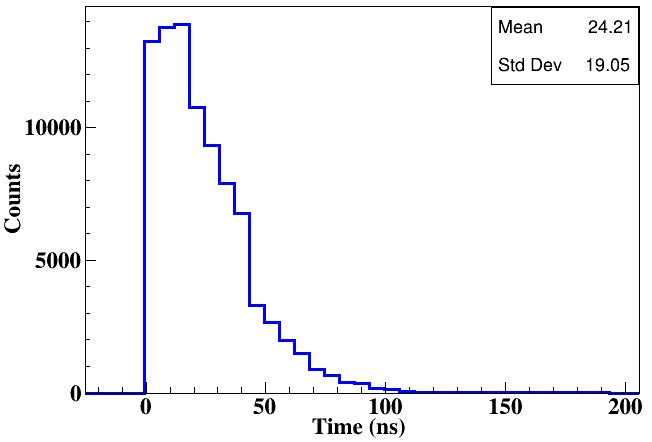}
  \vspace{-2mm}
  \caption{\label{fig:ClusterSizeTimeDiffAndADC} Left: Cluster size distribution for the entire GEM1 plane. 
  Middle: Cluster charge (ADC) distribution for GEM1. The distribution is fitted with a Landau distribution. 
  Right: Time separation of digis within clusters. The standard deviation yields to $\sim$19\,ns. }    
\end{figure}

The X (cm) - Y (cm) distribution of hits for GEM1 (left) and GEM2 (right) are shown in Fig.~\ref{fig:XYDistHitsG1G2}, revealing the detector acceptance. Figure~\ref{fig:ClusterSizeTimeDiffAndADC} describes the cluster characteristics for the hits in GEM1. The left panel shows the typical cluster size distribution for GEM1 at an operating voltage of $\Delta V_{GEM}(sum)$ $\sim$ 1072\,V. The corresponding distribution of the cluster charge (fC) for the entire GEM1 clusters is shown in the middle panel. Since we have used 5-bit ADC, a considerable charge falls in the overflow bin (31 ADC channel), hence a spike at around 82~fC. Similar nature of the cluster charge (fC) distribution has also been seen in response simulations (GEANT). The cluster charge spectra are fitted with a Landau distribution, giving an MPV of 15.13~fC. Using the MPV value, the gain at a summed GEM voltage of about 1072~V is estimated to be $\sim$3.1~$\times$10$^{3}$. Within a reconstructed cluster, the distribution of time-difference of all the digis with respect to the first Digi in time is shown in Fig.~\ref{fig:ClusterSizeTimeDiffAndADC} (right). The RMS of this plot gives an idea about the time spread between the digis of a MuCh cluster. It is observed that for more than 90\% of times, the separation between the Digis is within 50\,ns. This information is useful for optimizing the time separation between clusters related to the CBM 4D tracking.

\begin{figure}[htbp]
\centering 
\includegraphics[width=6.0cm,height=5.0cm]{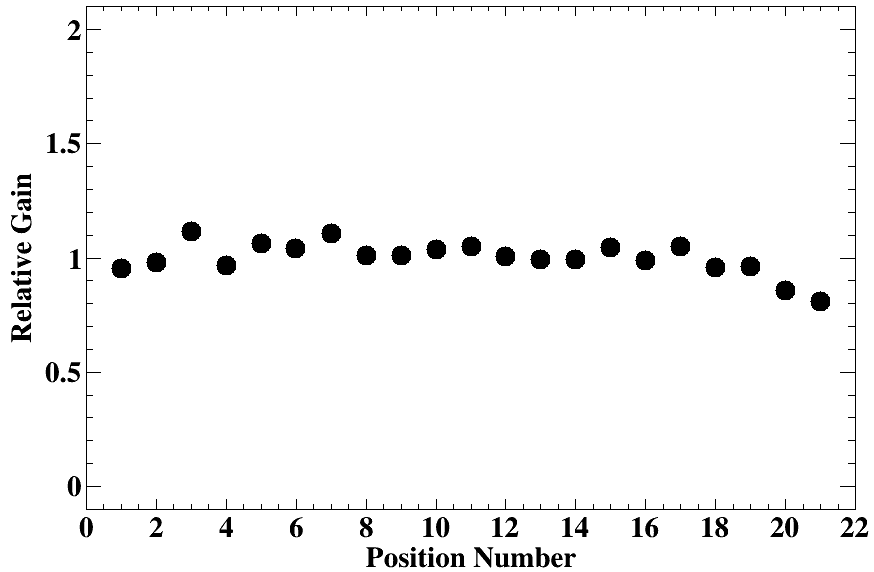}
\includegraphics[width=6.0cm,height=5.0cm]{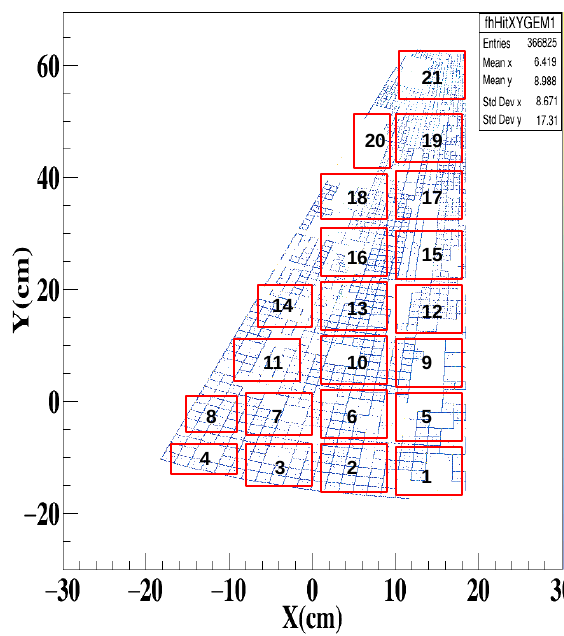}
\vspace{-4mm}
\caption{\label{fig:RelativeGain} Left: Relative gain of the detector in different regions for GEM1. 
Right: Position numbers used in the left plot is indicated on the detector plane.}
\end{figure}

\begin{figure}[htbp]
\centering 
\includegraphics[width=6.0cm,height=4.5cm]{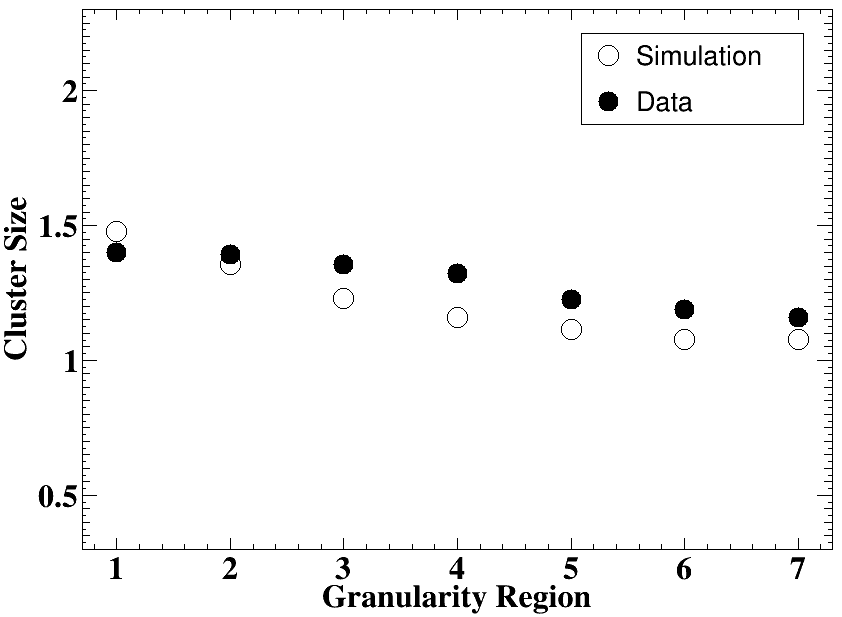}
\hspace{0mm}
\includegraphics[width=6.0cm,height=4.5cm]{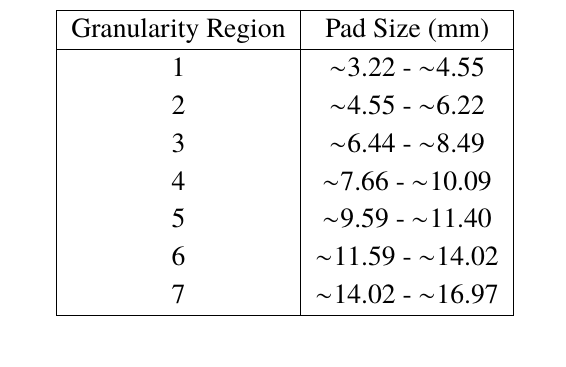}
\vspace{-3mm}
\caption{\label{fig:CluSizeWithGranularity} Left: Variation of cluster size at different granularity regions. 
Right: Table for granularity regions and their respective pad sizes.}
\end{figure}

The relative gain of the detector was measured at various positions on the detector plane, as shown in Fig.~\ref{fig:RelativeGain} (left). Each position number represents a specific zone in the detector, which is displayed in the right panel of the same figure. The gain distribution is observed to be uniform at the level of about 15\%. The readout plane has pads of varying granularity. The cluster size distribution in different regions of the detector have been measured. The variation of the average cluster size for different granularity zones is shown in Fig.~\ref{fig:CluSizeWithGranularity}. Statistical errors are included but within the marker size. Further contributions due to dead areas, noise, thresholds, fluctuations in gain, etc., are not considered. The different regions are indicated by numerals, and the adjoining table in the right indicate the corresponding granularity. The corresponding values obtained from GEANT simulations in the mCBM environment and realistic GEM module configurations are also overlaid, and the two values are found to match closely within 10\%.


\begin{figure}[htbp]
\centering 
\includegraphics[width=6.5cm,height=5.0cm]{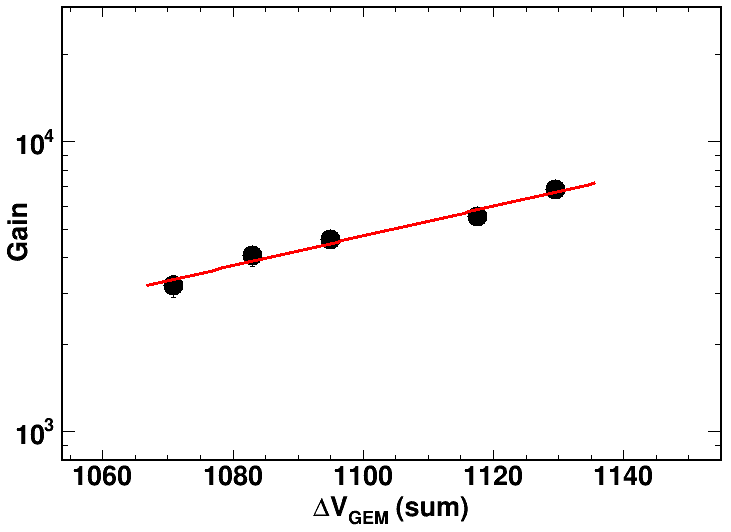}
\hspace{1mm}
\includegraphics[width=6.5cm,height=5.0cm]{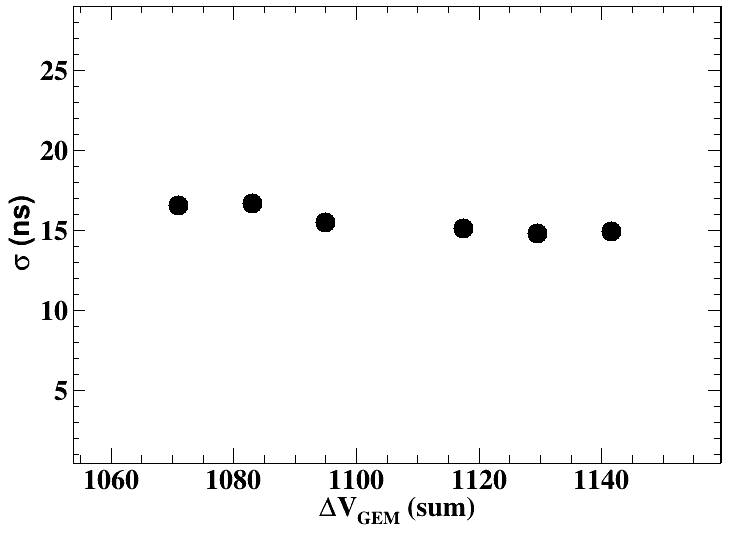}
\vspace{-4mm}
\caption{\label{fig:TimeResoAndGainVariationWithGEMV} The variation of gain (left) and time resolution (right) of the detector with the GEM voltages.}
\end{figure}

\begin{figure}[htbp]
\centering 
\includegraphics[width=6.5cm,height=5.0cm]{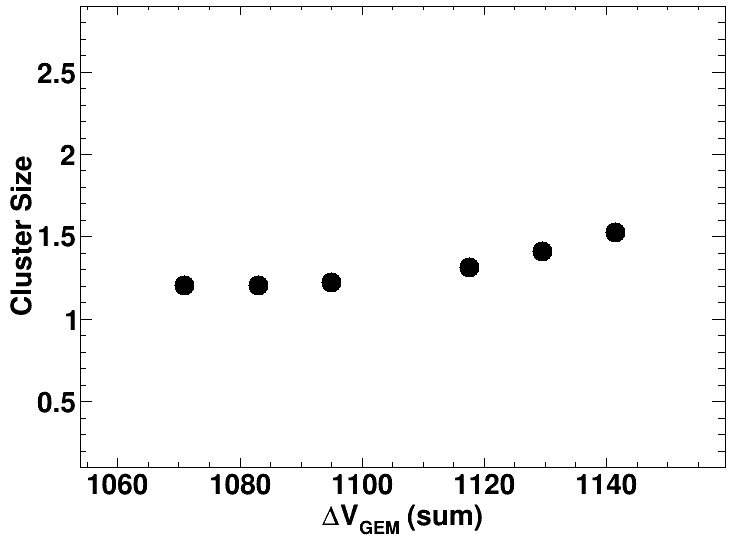}
\hspace{1mm}
\includegraphics[width=6.5cm,height=5.0cm]{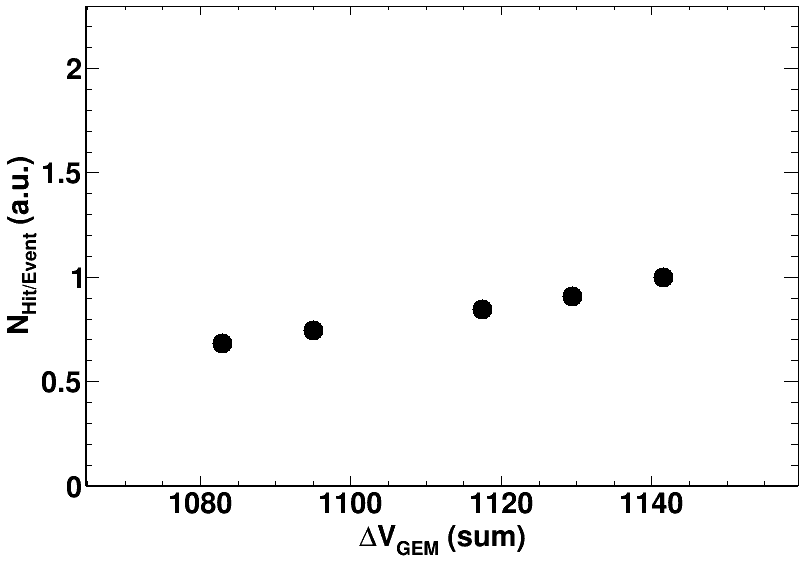}
\vspace{-4mm}
\caption{\label{fig:CluSizeAndHitVartaionWithGEMv}  Variation of cluster size with voltage (left). Variation of average Hit/event with GEM voltage.}
\end{figure}

Detector characteristics have been studied as a function of applied GEM voltages. The first among these is the variation of detector gain with the GEM voltage. As shown in Fig.~\ref{fig:TimeResoAndGainVariationWithGEMV} (left), the gain increases exponentially with increasing voltage, as expected. The right panel shows the variation of time resolution with applied voltage. The time resolution is observed to saturate at around 15\,ns. The errors (statistical) on the data points in both the panels are within the marker size. The variation of the cluster size with voltage is shown in Fig.~\ref{fig:CluSizeAndHitVartaionWithGEMv} (left). The slight increase, which is also observed in simulation, is due to increased gain at higher voltages. The variation of average hits per event is shown in Fig.~\ref{fig:CluSizeAndHitVartaionWithGEMv} (right). The curve shows a rising trend, indicative of the effect due to increased efficiency at higher GEM voltages.

\begin{figure}[htbp]
\centering 
\includegraphics[width=7.0cm,height=5.0cm]{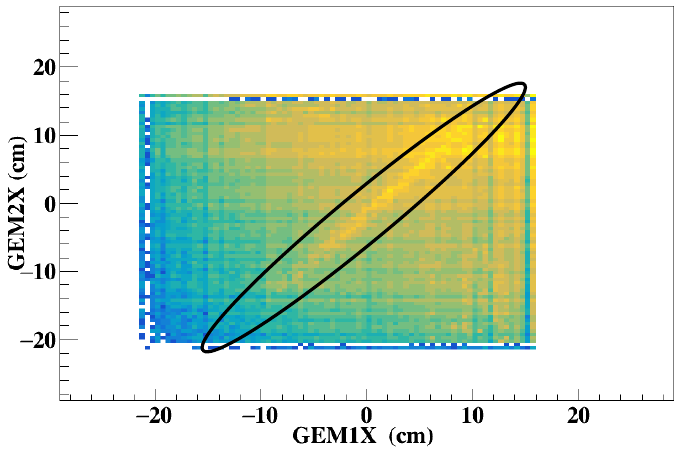}
\includegraphics[width=7.0cm,height=5.0cm]{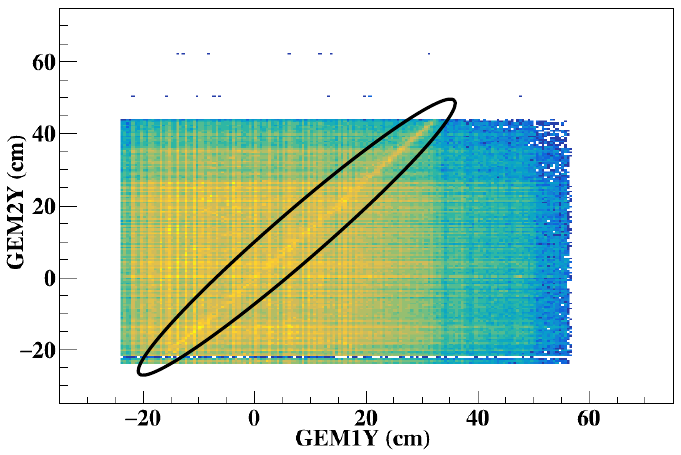}
\vspace{-4mm}
\caption{\label{fig:SpatialCorrG1G2} Spatial correlation in X (cm) and Y (cm) between GEM1 and GEM2 hits}
\end{figure}

Using the hits registered on the different detector planes within the built events described in the above section, we try to investigate the spatial correlation. Figure~\ref{fig:SpatialCorrG1G2} displays the spatial correlation in X and Y between GEM1 and GEM2. The range of the correlation line observed in both X and Y matches with the observations of the 2D-Hits plot of Fig.~\ref{fig:XYDistHitsG1G2}. The region of common overlap in X and Y is highlighted in the elliptical ring. It must be stated that the event building is required to make the correlation lines cleanly visible, which is not feasible while scanning complete time-slices of 10\,ms overlaying many events.

Further studies related to extracting detector efficiencies, multiplicity distributions, and the performance at high collision rates are expected to be carried out in the upcoming mCBM campaigns.

\section{Summary}
\label{sec:Summary}
We have commissioned large-size trapezoidal GEM modules designed for the CBM-MuCh system in the ongoing mCBM experiment at GSI. The modules have been tested under realistic experimental conditions in nucleus-nucleus collisions. Data have been taken in November and December 2019 for Ar~+~Au collisions at 1.7\,AGeV beam kinetic energy. A preliminary CBM data analysis chain based on the CbmRoot software framework has been applied to the data. First results have been obtained by studying the detector performance in terms of on-spill and off-spill count rates, time correlations and resolution.

Noisy channels were identified using the off-spill part of the data. Appropriate offset corrections have been implemented to enable a first event building by applying a time-cluster search. For the first time, a cluster and hit reconstruction has been performed on the MuCh detector data taken with the free-streaming CBM DAQ system. The cluster-size and cluster-charge characteristics in different zones of the detector have been studied as well. The gain of the detector has been calculated using the cluster-charge of reconstructed hits and its variation with the GEM voltage. An average time resolution of about 15\,ns was measured for the tested GEM module. The uniformity of the detector response has been studied in terms of the relative gain-plot using the mean of the cluster charge distribution zone-wise and in terms of the time resolution measured for a large number of pads throughout the detector area. An RMS of 4-5~ns describes the spread in the time resolutions measured over the entire area of the detector. The average dispersion of the digis in time within a cluster has been studied as well. Performing a first event building on digi-level, clear spatial correlations between both GEM modules (GEM1 and GEM2) were observed.
 
\acknowledgments
AK acknowledges the receipt of a DAE-HBNI Fellowship. The work has been funded by DAE, India. The results presented here are based on the experiment S471 (mCBM), which was performed at the beam line/infrastructure HTD at the GSI Helmholtzzentrum fuer Schwerionenforschung, Darmstadt (Germany) in the frame of FAIR Phase-0. We acknowledge the GSI team for the intensive help and fruitful discussions.


\bibliographystyle{unsrt}
\bibliography{mybib}

\end{document}